# Four-wave mixing and nonlinear parameter measurement in a gallium-nitride ridge waveguide

DVIR MUNK,[1] MOSHE KATZMAN,[1] OHAD WESTREICH,[2,3] MORAN BIN NUN,[3] YEDIDYA LIOR,[3] NOAM SICRON,[2] YOSSI PALTIEL,[3] AND AVI ZADOK[1,*]

[1]*Faculty of Engineering and Institute for Nano-Technology and Advanced Materials, Bar-Ilan University, Ramat-Gan 5290002, Israel*
[2]*Solid State Physics Department, Applied Physics Division, Soreq NRC, Yavne 81800, Israel*
[3]*Applied Physics Department, Hebrew University, Jerusalem 91904, Israel*
*\*Avinoam.Zadok@biu.ac.il*

**Abstract:** Gallium-nitride (GaN) is a promising material platform for integrated electro-optic devices due to its wide direct bandgap, pronounced nonlinearities and high optical damage threshold. Low-loss ridge waveguides in GaN layers were recently demonstrated. In this work we provide a first report of four-wave mixing in a GaN waveguide at telecommunication wavelengths, and observe comparatively high nonlinear propagation parameters. The nonlinear coefficient of the waveguide is measured as 1.6±0.45 [W×m]$^{-1}$, and the corresponding third-order nonlinear parameter of GaN is estimated as 3.4±1e-18 [m$^2$/W]. The results suggest that GaN waveguides could be instrumental in nonlinear-optical signal processing applications.

## 1. Introduction

Gallium-nitride (GaN) is an excellent candidate material for many electro-optic and integrated-photonic devices. The wide, direct bandgap of GaN is being used in blue light-emitting diodes and laser diodes [1,2], and in detectors for blue and ultraviolet light [3]. Due to its non-centro-symmetric crystalline structure, GaN exhibits piezo-electricity [4] and second-order nonlinear optical effects [5,6]. GaN is also characterized by a low thermo-optic coefficient, weak material dispersion, high optical damage threshold and chemical stability, and it is suitable for operation at high temperatures and for hybrid integration on silicon [7].

Motivated by the above advantages, waveguides and light-guide circuits in thin GaN films are drawing increasing attention [8-14]. In particular, the absence of two-photon and even three-photon absorption at telecommunication wavelengths holds much promise for the use of GaN waveguides in all-optical nonlinear signal processing applications. Such functions were successfully demonstrated in other integrated photonics platforms, such as chalcogenide glasses [15] or silicon nitride [16].

Third-order nonlinear parameters of GaN films were investigated in several works using Z-scan methods [17-21]. Most attention was given to characterization at visible or near-infrared wavelengths, where the extracted nonlinear coefficients are often wavelength-dependent and include contributions of two-photon absorption and free carriers effects. Third-order nonlinear parameters of GaN at 1550 nm wavelength were not measured directly. The extrapolation of data taken at shorter near-infrared wavelengths suggests values on the order of 1e-18 [$m^2$/W] [22,23]. Studies of third-order nonlinear propagation effects in GaN waveguides are few and far between. In some cases, the nonlinear propagation was restricted by comparatively high propagation losses. In one recent example, third harmonic generation was demonstrated in a photonic crystal GaN waveguide [24].

Recently, Westreich *et al.* demonstrated ridge waveguides in GaN with propagation losses as low as 1 dB/cm at 1550 nm [25]. These low losses make nonlinear-optical signal processing a realistic prospect. In this work, we provide a first report of four-wave mixing (FWM) in GaN ridge waveguides at 1550 nm. The device under test was only 3.5 mm long, yet the measurement setup was sensitive enough to provide quantitative estimates of nonlinear propagation parameters. The results suggest that the third-order nonlinear parameter of FWM in GaN at 1550 nm is comparable with that of $As_2S_3$ chalcogenide glass [26], and an order of magnitude larger than that of silicon nitride [27]. The results demonstrate that GaN waveguides may provide a useful platform for nonlinear optics applications.

The remainder of this paper is organized as follows: The design and fabrication of the GaN waveguides under test are presented in section 2, FWM measurements are reported in section 3, and a brief concluding discussion is given in section 4.

## 2. Design and fabrication of GaN ridge waveguides

Previous works present a detailed report of the design, fabrication and linear characterization of the GaN waveguides [25,28]. A brief description is provided here for completeness. Waveguides were fabricated in GaN / AlGaN layers grown on a c-plane sapphire substrate, using metal-organic chemical vapor deposition. The layers structure is illustrated schematically in Fig. 1(a). A 1.5 µm-thick core layer of GaN was grown between lower and upper cladding layers of $Al_{0.03}Ga_{0.97}N$, with respective thicknesses of 0.3 µm and 0.8 µm. Ridge waveguides of different widths were defined using a $SiO_2$ hard mask and standard lithography, followed by inductively coupled plasma reactive ion etching. Ridges were etched to a depth of 1.0 µm. A 200 nm-thick layer of SiO$_2$ was deposited on top the waveguides to reduce propagation losses. The waveguides were diced and their facets were polished using suspensions of diamond and silica particles. Figure 1(b) shows the numerically-calculated transverse profile of the single transverse-electric (TE) mode that is supported by a 4 µm-wide waveguide. A scanning-electron microscope cross-section image of the facet of a 10 µm-wide ridge waveguide is shown in Fig. 1(c). The effective index of the TE mode at 1550 nm was calculated as 2.26 RIU. That index is within 0.02 RIU of the refractive index of bulk GaN at 1550 nm and the same polarization [29]. The relative confinement of optical power to the GaN layer is 84%. Additional 15% are confined to the Al$_{0.03}$Ga$_{0.97}$N cladding layers.

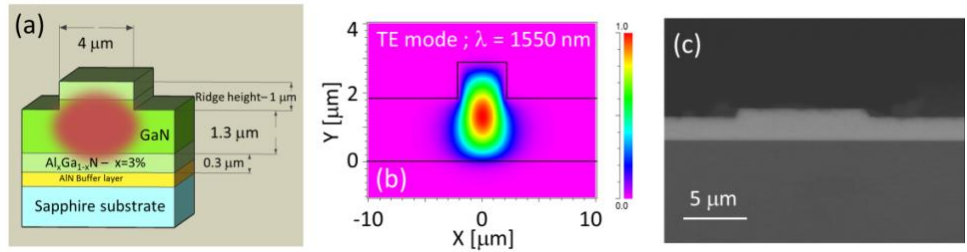

Fig. 1. (a): Schematic illustration of a 4 µm-wide GaN ridge waveguide, presenting a cross section of the layers structure. (b): Calculated transverse profile of the fundamental transverse-electric mode supported by a 4 µm-wide GaN ridge waveguide. (c): Scanning electron microscope image of the facet of a 10 µm-wide GaN ridge waveguide.

### 3. Four-wave mixing and nonlinear index measurements

The experimental setup used in the demonstration of FWM in a GaN ridge waveguide is illustrated in Fig. 2. The outputs of two laser diodes of wavelengths $\lambda_1 = 1549.9$ nm (frequency $\omega_1$) and $\lambda_2 = 1549.4$ nm (frequency $\omega_2$) were used as FWM pump waves. Light from each diode passed through a separate electro-optic amplitude modulator (EOM, see also below). Both modulators were driven by pulses of 10-60 ns duration and 1 μs period. The two pump waves were then amplified by erbium-doped fiber amplifiers (EDFAs), and coupled together by a 50/50 fiber coupler. Modulation by low-duty-cycle pulses provided high peak power levels at the output of the EDFAs, and helped enhance FWM accordingly. The pattern generators driving the two EOMs were synchronized so that the pulses at both wavelengths overlapped in time. A tunable narrowband optical bandpass filter (BPF) was adjusted to transmit the two pump waves, and reject any background light from the laser diodes or EDFAs at the wavelengths of expected FWM products. The average power of each pump wave at the BPF output was 15.5 dBm. The peak power levels $P_{1,2}$ of pump pulses at wavelengths $\lambda_{1,2}$ increased for shorter pulses, and were monitored for each pulse duration. The two levels were kept equal to within 1 dB.

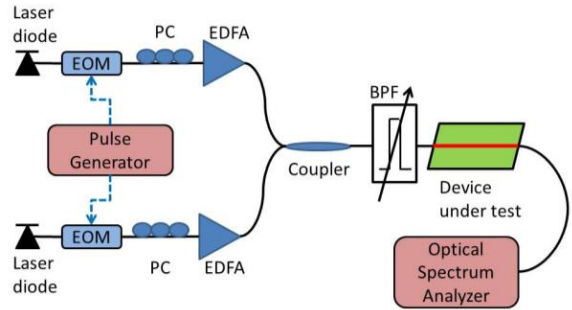

Fig. 2. Illustration of the setup used in four-wave mixing measurements in a GaN waveguide. EOM: electro-optic modulator. EDFA: erbium-doped fiber amplifier. PC: polarization controller. BPF: optical bandpass filter.

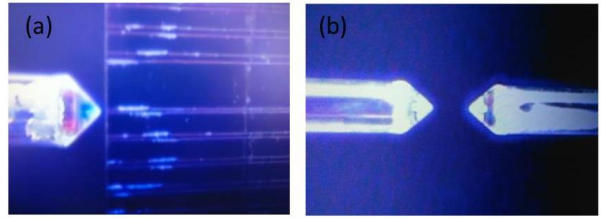

Fig. 3. (a): Top-view image of GaN waveguides alongside a lensed fiber. (b): Input and output lensed fibers coupled directly in a reference measurement, bypassing the waveguide under test.

The pump waves were coupled into a GaN ridge waveguide using a lensed fiber with spot-size diameter of 2 µm. The waveguide under test was 3.5 mm long, with a ridge width of 4 µm and an effective modal area $A_{eff}$ of 9 µm². Figure 3(a) shows a top-view image of several waveguides alongside the lensed fiber. Polarization controllers (PCs) were used to adjust the states of polarization of both pump waves for maximum end-to-end transmission through the waveguide, a condition which corresponds to TE alignment [25]. Light at the output end of the waveguide was collected by a second, identical lensed fiber. The end-to-end transmission loss of optical power was measured as -9.4 dB. Based on previous characterization [25], propagation losses within the short waveguide are estimated as 0.6 dB. We assume hereunder input and output power coupling efficiencies $\eta_{in} \approx \eta_{out} \approx$ -4.4 dB. FWM terms at frequencies $\omega_3 = 2\omega_1 - \omega_2$ ($\lambda_3$ = 1550.4 nm) and $\omega_4 = 2\omega_2 - \omega_1$ ($\lambda_4$ = 1548.9 nm) were measured using an optical spectrum analyzer.

Although the third-order nonlinearity in the GaN waveguide is orders of magnitude larger than that of single-mode fiber, the accumulation of FWM over many meters of fiber could be much stronger than the corresponding effect in a millimeter-scale device. The two pump waves were therefore modulated and amplified in separate fiber paths, in attempt to reduce their co-propagation in a single fiber to a minimum. Any FWM along the fiber path between the coupler and the BPF was effectively blocked by the BPF. However, the output path of the BPF still contained several meters of fibers which could not be removed. The length of this lead-in fiber could not be measured directly. The contribution of FWM along the lead-in fiber was therefore taken into consideration in the analysis of experimental data. The peak optical power levels of the FWM products at wavelengths $\lambda_{3,4}$ at the device output are given by:

$$P_3 = \eta_{in}\eta_{out} P_1^2 P_2 \left(\gamma_F L_F + \eta_{in}\gamma_{WG} L_{WG}\right)^2, \tag{1}$$

$$P_4 = \eta_{in}\eta_{out} P_2^2 P_1 \left(\gamma_F L_F + \eta_{in}\gamma_{WG} L_{WG}\right)^2. \tag{2}$$

Here $\gamma_{F,WG}$ are the nonlinear coefficients of the lead-in fiber and waveguide respectively, in units of [W×m]$^{-1}$, and $L_{F,WG}$ stand for the lengths of the fiber and waveguide. The effective index of the TE mode of the 4 µm-wide waveguide was calculated for $\{\lambda_i\}$, $i=1...4$, based on the design geometry and the extrapolation of material dispersion values given in the literature [29]. The variations in effective index among the four wavelengths are on the order of 6e-5 RIU. The corresponding phase mismatch for the FWM process over a 3.5 mm-long device is 7 mrad. This mismatch has a negligible effect on the FWM efficiency.

The FWM contributions of fiber and waveguide were separated using the following procedure: First the device under test was bypassed, and the input and output lensed fibers were located facing each other directly, at a carefully controlled distance. The gap between the two fibers was adjusted to obtain direct coupling efficiency that matched the end-to-end transmission through the device. FWM products observed in the output spectrum were generated along the input fiber path only, with optical powers of $P_3 = \eta_{in}\eta_{out} P_1^2 P_2 \left(\gamma_F L_F\right)^2$ and $P_4 = \eta_{in}\eta_{out} P_2^2 P_1 \left(\gamma_F L_F\right)^2$. These measurements provided the necessary calibration of the term $\gamma_F L_F$. A second measurement of FWM power was then taken with the pump waves coupled through the device. Given the prior measurement of $\gamma_F L_F$, the nonlinear coefficient of the waveguide $\gamma_{WG}$ could be determined using Eq. (1) and Eq. (2).

Figure 4(a) shows examples of the power spectra of light at the waveguide output (blue trace), and with the waveguide bypassed as discussed above (red trace). The duration of pump pulses was 30 ns, and the peak power level of each pump pulse at the BPF output was 30.2±0.5 dBm. FWM terms at $\lambda_{3,4}$ are observed in both traces. The FWM products at the waveguide output are stronger than those of the reference trace. Figure 4(b) shows the FWM efficiencies of both FWM terms as a function of the peak power levels of the pump pulses. The efficiencies obtained at the waveguide output are consistently higher than those in the input fiber path alone. The efficiencies scale with the peak pump power squared, as expected. Differences between the efficiencies at $\lambda_{3,4}$ are due to residual differences between the peak power levels of the two pump waves.

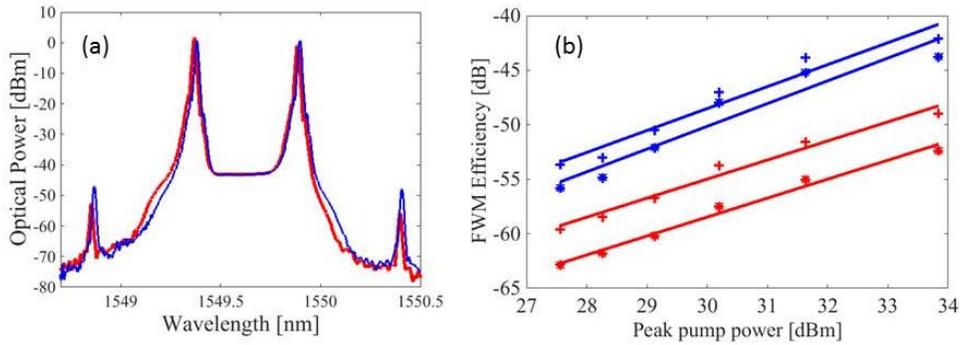

Fig. 4. (a): Measured optical power spectral densities at the output of a GaN ridge waveguide (blue), and with the waveguide bypassed (red). Pump pulses were of 30 ns duration and 1 μs period. The peak power level of each pump pulse was 30.2 dBm. Four-wave mixing terms to the sides of the two incident pump waves are observed. The process efficiency at the waveguide output is higher than that of the reference trace. (b): Four-wave mixing efficiencies (dB) as a function of the peak power level of each pump pulse (dBm). Blue markers and lines represents measurements at the waveguide output, whereas red markers and lines correspond to reference measurements with the waveguide bypassed. Asterisk markers: four-wave mixing products at $\lambda_3 =$ 1550.4 nm. Plus markers: four-wave mixing products at $\lambda_4 = $ 1548.9 nm. Solid lines show linear fits for the log-scale efficiencies as a function of log-scale peak power. The fitting slopes are 1.9 ± 0.15, in agreement with the expected square-law dependence.

The measured FWM efficiencies correspond to a nonlinear coefficient $\gamma_{WG}$ of 1.6±0.3 [W×m]$^{-1}$. This value is 1,200 times larger than known values of $\gamma_F$. The experimental uncertainty in the input coupling efficiency $\eta_{in}$ is estimated as ±1 dB. This uncertainty corresponds to an additional, statistically independent experimental error of ±0.3 [W×m]$^{-1}$ in the nonlinear coefficient $\gamma_{WG}$. The overall experimental uncertainty in the measurement of $\gamma_{WG}$ is therefore ±0.45 [W×m]$^{-1}$. Lastly, the third-order nonlinear parameter of GaN corresponding to FWM at 1550 nm was estimated according to: $\gamma_{WG} A_{eff}/k_0$, where $k_0$ is the vacuum wavenumber. A value of 3.4±1e-18 [m$^2$/W] was obtained. Since the optical mode is largely confined to the GaN core layer, the estimated nonlinear parameter provides a good measure for the GaN material property.

### 4. Conclusions

A first demonstration of FWM in a ridge GaN waveguide has been provided. The nonlinear coefficient of a short device under test could be resolved with careful calibration of FWM contributions from the fiber setup. The measurements provided an estimate for the nonlinear third-order parameter of FWM in GaN at 1550 nm, on the order of 3.4e-18 [m$^2$/W]. This estimate is consistent with previous data that is extrapolated from Z-scan measurements of GaN films at shorter, near-infrared wavelengths [22,23]. The estimated value is two orders of magnitude larger than that of silica, an order of magnitude larger than that of silicon nitride [27], and similar to that of As$_2$S$_3$ [26]. The corresponding nonlinearity of silicon is larger than that of GaN [30]. However, in contrast to silicon, nonlinear propagation in GaN at 1550 nm wavelength is not restricted by two-photon absorption.

The results suggest that GaN ridge waveguides constitute a very useful platform for nonlinear-optical signal processing. Low propagation losses may provide effective lengths of several cm. The high optical damage threshold of GaN and the absence of two-photon and three-photon absorption at 1550 nm imply that high intensities can be coupled into devices. The availability of the piezo-electric and electro-optic effects alongside the relatively high nonlinear index opens up additional possibilities for realizing active photonic devices [4-6]. Lastly, the integration of GaN devices alongside silicon photonics may enhance the nonlinear processing capabilities of the latter platform. Such integration has already been proposed in the literature [31], using wafer bonding techniques. Coupling between the layers may be controlled through the thickness of intermediate oxide layers. Future works would attempt to demonstrate more efficient nonlinear processing over longer GaN waveguides, and explore the potential of this platform further.

**Funding**
European Research Council (ERC), Starter Grant H2020-ERC-2015-STG 679228 (L-SID).

**Acknowledgments**
The authors thank Dr. Arkady (Arik) Bergman of Bar-Ilan University for assistance with the experimental setup. D.M., M.K. and A.Z. acknowledge the support of the European Research Council (ERC), through Starter Grant H2020-ERC-2015-STG 679228 (L-SID).